\documentclass[twocolappendix]{emulateapj}
\usepackage{apjfonts,natbib,graphicx,amsmath,xspace, ulem}
\usepackage{hyperref}
\usepackage{xspace}
\graphicspath{{./}}

\newcommand{\pks}{PKS~2005$-$489\xspace}
\newcommand{\mrk}{Mrk~501\xspace}
\newcommand{\fes}{1ES~1101$-$232\xspace}
\def\s5{S5~$0716+714$\xspace}
\newcommand{\xte}{\textit{RXTE}\xspace}

\def\lsim{\mathrel{\rlap{\lower4pt\hbox{\hskip1pt$\sim$}}
    \raise1pt\hbox{$<$}}}                
\def\gsim{\mathrel{\rlap{\lower4pt\hbox{\hskip1pt$\sim$}}
    \raise1pt\hbox{$>$}}}                

\begin{document}

\title{Extremely Rapid X-ray flares of TeV Blazars in the \xte{} Era}

\author{S.~F.~Zhu\altaffilmark{1,2,3}, Y.~Q.~Xue\altaffilmark{2,3}, W.~N.~Brandt\altaffilmark{1,4,5}, W.~Cui,\altaffilmark{6} and Y.~J.~Wang\altaffilmark{2,3}}

\altaffiltext{1}{Department of Astronomy \& Astrophysics, The Pennsylvania State University, University Park, PA 16802; sxz89@psu.edu}
\altaffiltext{2}{CAS Key Laboratory for Research in Galaxies and Cosmology, Department of Astronomy, University of Science and Technology of China, Hefei 230026, China; xuey@ustc.edu.cn}
\altaffiltext{3}{School of Astronomy and Space Science, University of Science and Technology of China, Hefei 230026, China}
\altaffiltext{4}{Institute for Gravitation and the Cosmos, The Pennsylvania State University, University Park, PA 16802}
\altaffiltext{5}{Department of Physics, 104 Davey Lab, The Pennsylvania State University, University Park, PA 16802}
\altaffiltext{6}{Department of Physics and Astronomy, Purdue University, West Lafayette, IN 47907}

\begin{abstract}
Rapid flares from blazars in very high energy (VHE) $\gamma$-rays challenge the common understanding of jets of active galactic nuclei (AGNs).
The same population of ultra-relativistic electrons is often thought to be responsible for both X-ray and VHE emission.
We thus systematically searched for X-ray flares at sub-hour timescales of TeV blazars in the entire \textit{Rossi X-ray Timing Explorer} archival database.
We found rapid flares from \pks{} and \s5{}, and a candidate rapid flare from \fes{}.
In particular, the characteristic rise timescale of \pks{} is less than half a minute,
which, to our knowledge, is the shortest among known AGN flares at any wavelengths.
The timescales of these rapid flares indicate that the size of the central supermassive black hole is not a hard lower limit on the physical size of
the emission region of the flare.
\pks{} shows possible hard lags in its flare, which could be attributed to particle acceleration (injection);
its flaring component has the hardest spectrum when it first appears.
For all flares, the flaring components show similar hard spectra with $\Gamma=1.7-1.9$,
and we estimate the magnetic field strength $B\sim$ 0.1--1.0 G by assuming synchrotron cooling.
These flares could be caused by inhomogeneity of the jets.
Models that can only produce rapid $\gamma$-ray flares but little synchrotron activity are less favorable.

\end{abstract}

\section{Introduction}

Blazars, including BL Lac objects and flat-spectrum radio quasars (FSRQs),
are a special class of radio-loud active galactic nuclei (AGNs) that have one of their relativistic jets pointing very close to our line of sight \citep[e.g.,][]{Urry1995}.
FSRQs have luminous broad emission lines that are weak or absent in BL Lac objects.
Due to Doppler boosting, the emission of a blazar is usually dominated by the jet whose spectral energy distribution (SED) shows two broad humps that smoothly extend from radio to $\gamma$-rays.
The low-energy hump can extend from radio to soft X-rays.
According to the frequency of the first hump,
BL Lac objects are further divided into low-frequency peaked BL Lac objects (LBLs; $\nu_{\rm peak}<10^{14}$ Hz),
intermediate-frequency peaked BL Lac objects (IBLs; $10^{14}<\nu_{\rm peak}<10^{15}$ Hz) and high-frequency peaked BL Lac objects (HBLs; $\nu_{\rm peak}>10^{15}$ Hz) \citep[e.g.,][]{Padovani1995,Abdo2010a}.
The high-energy hump extends from hard X-rays to $\gamma$-rays, even sometimes the very high energy (VHE) TeV band.
Such VHE blazars, typically HBLs, are called TeV blazars.

The low-energy hump is attributed to the synchrotron emission of highly relativistic electrons gyrating in a magnetic field in the jet.
The origin of the high-energy hump, however, is still debated.
A popular explanation is inverse-Compton emission from the same population of relativistic electrons that produce the synchrotron emission.
The seed photons of the inverse-Compton scattering process could be local synchrotron photons (usually for BL Lacs)
in the jet and/or external photons from the central engine (usually for FSRQs) or the cosmic microwave background.
These models are thus called leptonic models. The hadronic models, on the other hand,
attribute $\gamma$-ray emission to synchrotron emission of protons \citep{Mucke2001, Mucke2003, Fraija2015a} or proton-induced cascades \citep{Mannheim1998}.

The observed photon flux of blazars varies significantly across the electromagnetic spectrum on timescales from minutes to years \citep[e.g.,][]{Wagner1995,Ulrich1997}.
The origin of the variability is not well understood.
Generally, the variability is noise-like \citep[e.g.,][]{Kataoka2001, Chatterjee2012}, similar to the variability
of radio-quiet AGNs \citep[e.g.,][]{Markowitz2003}. However, blazars are also known to have bursts that show flare-like structures \citep[e.g.][]{Marscher2010}, which
may have recognizable patterns \citep{Sasada2017}. The outbursts can be explained by internal shocks of the jets when a new relativistic
blob of plasma catches up with an old blob and accelerates particles to ultra-relativistic energies \citep[e.g.,][]{Spada2001}.
Based on several similarities between the jet emission and corona-disk emission,
the ultimate origin of the variability may still
be accretion-rate fluctuations of the disk \citep{McHardy2008}.

The shortest variability timescale is a crucial parameter because it serves as an independent constraint on the physical scale of the emission region \citep{Tavecchio1998}, which cannot
be easily provided by other observational measurements.
Blazars are usually most variable at frequencies just above the two SED humps \citep[e.g.,][]{Ulrich1997, Madejski2016},
which usually fall in the hard X-ray and TeV bands in the case of TeV blazars \citep[e.g.,][]{Aleksic2015a, Balokovic2016, Bartoli2016}.
In particular, an increasing number of TeV blazars show $\gamma$-ray flaring activity on timescales from several to a few tens of minutes
that are detected by ground-based Cherenkov telescopes,
including both BL Lac objects \citep[e.g.,][]{Gaidos1996, Aharonian2007, Albert2007, Arlen2013} and FSRQs \citep{Aleksic2011}.
The minute-scale variability in the TeV band \citep[e.g.][]{Aharonian2007} has strong implications for our understanding of AGN jets \citep{Begelman2008}.
X-ray and TeV emission may be directly related to the same high-energy tail of the relativistic electron population.
Indeed, the lightcurves of HBL-type TeV blazars in the X-ray and TeV bands are usually correlated \citep[e.g.,][]{Aleksic2015b, Furniss2015}.
Attempts to search for extremely rapid X-ray variability have been made \citep[e.g.,][]{Cui2004, Xue2005, Pryal2015, Paliya2015}.
The same source can have minute-scale variability in both the X-ray band and TeV band \citep[e.g., Mrk501,][]{Xue2005, Albert2007}.
However, ``orphan'' TeV flares that have no X-ray counterparts are occasionally reported \citep[e.g.,][]{Krawczynski2004, Blazejowski2005, Acciari2009, Fraija2015b}.

Rapid TeV variability has germinated various models to explain the small timescales.
Most models involve some very compact regions moving in the rest frame of the jet.
These compact regions could be ``jets in a jet''
that are either produced by magnetic reconnection processes in a Poynting flux-dominated jet \citep{Giannios2009}
or relativistic turbulence in the jet \citep{Narayan2012}.
The minijets-in-a-jet model can consistently produce the statistical properties of blazar flux \citep{Biteau2012}.
There are also models involving a red giant star being stripped of its envelope by the jet \citep{Barkov2012}
and models involving beams of magneto-centrifugally accelerated electrons occasionally pointing toward us \citep{Ghisellini2009}.

Several well-studied TeV blazars show rich spectral behavior in X-rays, which may represent
the general behavior of the synchrotron peak of all AGN jets.
The X-ray spectra are usually curved \citep{Massaro2004} and can only locally be fitted by a power-law.
The spectral variation with flux can be complex \citep{Zhang2002, Cui2004}.
Generally, the spectrum hardens when the flux increases \citep[e.g.,][]{Gliozzi2006,Xue2006,Tramacere2009},
but photon indexes can saturate at higher fluxes \citep{Xue2005, Giebels2007}.
The synchrotron peak usually moves to higher frequencies with increasing flux during outbursts \citep[e.g.][]{Pian1998},
but no correlation between the break energy and the flux exists when a broken power law is adopted to fit the X-ray spectra \citep{Xue2005, Giebels2007, Garson2010}.
A cooling break in the spectrum of emitting particles cannot explain these features \citep{Wierzcholska2016b}, and
some special particle acceleration processes may be involved \citep{Madejski2016}.
There are also energy-dependent lags between the variations of different energy bands.
In some flares, soft bands lag behind hard bands \citep[e.g.,][]{Zhang2002}, while lags in the opposite direction can also happen \citep[e.g.,][]{Ravasio2004, Sato2008}.
Hysteresis in the HR (hardness ratio)--flux diagram is often used as a diagnostic of lags.
Clockwise loops \citep[e.g.,][]{Acciari2009, Kapanadze2016} in the HR--flux plane are a sign of soft lags
while counterclockwise loops \citep[e.g.][]{Tramacere2009} are a sign of hard lags.
The same source can exhibit both clockwise and counterclockwise loops;
the observed patterns are further complicated by the superposition of flares at different timescales \citep{Cui2004}.
The above knowledge of TeV blazars in the X-ray regime comes from studies focusing on timescales of hours to weeks.
We will extend this kind of analysis to much smaller timescales in this paper.

The main goal of this paper is to search for X-ray flares at sub-hour timescales from TeV blazars
in the entire \textit{Rossi X-ray Timing Explorer} ({\it RXTE}) archival database.
We use data from the narrow-field pointing instrument Proportional Counter Array (PCA) onboard {\it RXTE},
covering a nominal energy range of 2--60 keV.
The {\it RXTE} satellite was launched in December 1995 and ceased science operation in January 2012.
During its lifetime, it accumulated more than $\sim16$ Ms of exposure time on TeV blazars in hard X-rays, surpassing any other X-ray observatory.
We describe data reduction and the searching results in Section~\ref{sec:reduction}.
Most of our following analysis is based on an assumption that the observed photons are from
a flaring component and an underlying constant/slowly-varying component, possibly from two separated sites.
We describe lightcurve-model fitting and spectral-model fitting in Section~\ref{sec:lc} and Section~\ref{sec:spec}, respectively.
We discuss the implications of our findings in Section~\ref{sec:discussion} and summarize them in Section~\ref{sec:summary}.
In the following, we use the $\Lambda$CDM model, with $H_0=67.7$ km/s/Mpc and $\Omega_{\rm m}=0.307$ \citep{Planck2016}.

\section{Data reduction and searching for fast flares}
\label{sec:reduction}

\begin{table}[!t]
    \caption{TeV blazars with $>50$ {\it RXTE/PCA} observations}
    \centering
    \begin{tabular}{ccccc}\hline\hline
        Name & $z$ & Type & Number of & Exposure\\
        & & & Observations & Time (ks) \\
        \hline
        3C 279 & 0.5362 & FSRQ & 1988 & 3198\\
        BL Lacertae & 0.059 & IBL & 1387 & 2522 \\
        Mrk 421 & 0.031 & HBL & 1190 & 2515 \\
        PKS 1510$-$089 & 0.361 & FSRQ & 1334 & 2254 \\
        PKS 2155$-$304 & 0.116 & HBL & 501 & 1107 \\
        Mrk 501 & 0.034 & HBL & 499 & 886 \\
        S5 0716+714 & 0.31 & IBL & 233 & 733  \\
        H 1426+428 & 0.129 & HBL & 164 & 527 \\
        PKS 2005$-$489 & 0.071 & HBL &  158 & 483 \\
        3C 66A & 0.41 & IBL & 99 & 373 \\
        PKS 1424+240 & \nodata & HBL & 64 & 347 \\
        1ES 0229+200 & 0.14 & HBL & 205 & 295 \\
        1ES 1959+650 & 0.048 & HBL & 147 & 272 \\
        1ES 1101$-$232 & 0.186 & HBL & 99 & 211 \\
        1ES 2344+514 & 0.044 & HBL & 53 & 134 \\
        \hline
    \end{tabular}
    \begin{flushleft}
        {\sc Note.} --- We list above all the TeV blazars with $>50$ PCA observations that add up to $>130$ ks exposure time.
        The remaining unlisted TeV blazars have $<50$ PCA observations, which are PG~1553+113, 1ES~1218+304, MAGIC~J2001+435, 1ES~0806+524, 1ES~0647+250, RGB~J0152+017, 1ES~0414+009,
        W~Comae, 1ES~1727+502, Mrk~180, PKS~0447$-$439, RGB~J0710+591, PKS~0548$-$322, AP~Librae, 1ES~1741+196, and H~2356$-$309.
        See http://tevcat.uchicago.edu/ for the full list of known TeV blazars and their redshifts and classifications.
    \end{flushleft}
    \label{tab:srcList}
\end{table}

\begin{figure}[!t]
\centering
\includegraphics[width=0.45\textwidth, clip]{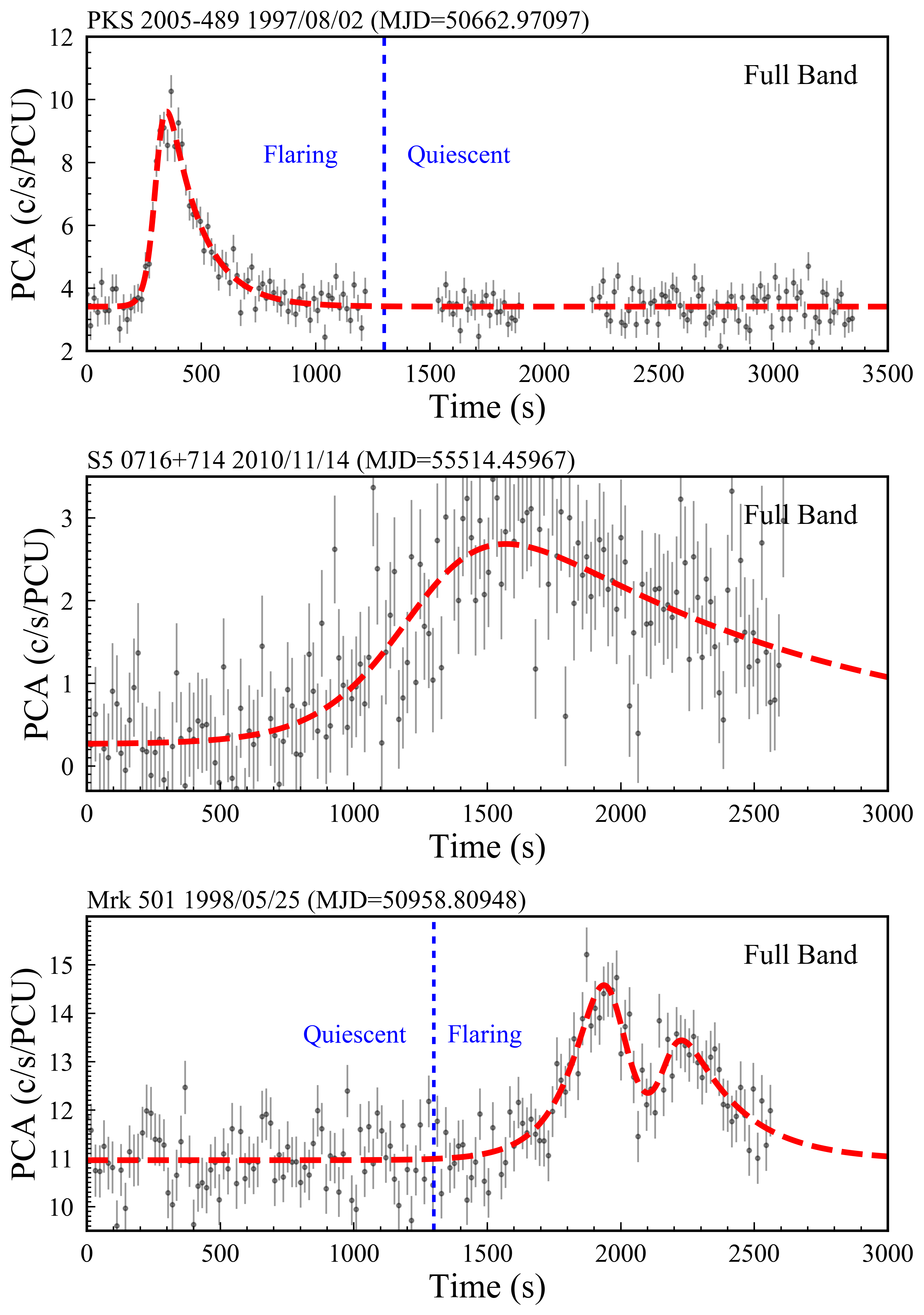}
\caption{Rapid X-ray flares of TeV blazars in bins of 16s. The red dashed curves are the weighted least square models (Eq.~\ref{eqn:flare_profile}).
The vertical dashed lines are used to separate the flaring phase from the quiescent phase in the light curve if possible.
    The event of \mrk{} was first reported by \cite{Xue2005}, which we reanalyzed more quantitatively in this paper.
The count rates for the $y$-axes are normalized to one PCU.
The corresponding ObsIDs of the three events from top to bottom are 20342-03-01-01, 95377-01-91-00, and 30249-01-01-02.
    On top of each panel, the date (Modified Julian Date) when the flaring observation started (i.e., set as $t=0$~s) is annotated.}
\label{fig:micro_flares}
\end{figure}

We retrieved all the archival {\it RXTE/PCA} observations\footnote{The data were downloaded from http://heasarc.gsfc.nasa.gov/cgi-bin/W3Browse/w3browse.pl.}
of TeV blazars (see Table~\ref{tab:srcList}). The total number of observations is $\sim8400$, and the total exposure time is $\sim16$ Ms.
We used the Standard 2 mode data, which have a time resolution of 16 s.
The data were reduced following the standard procedure.\footnote{See http://heasarc.gsfc.nasa.gov/docs/xte/recipes/cook\_book.html.}
We created filter files and good time intervals (GTIs) for each observation according to the suggested screening criteria\footnote{See ``Creating Filter Files and GTI Files for Use with Faint Models'' at http://heasarc.gsfc.nasa.gov/docs/xte/pca\_news.html.}
for faint sources.
Background data were then simulated using the appropriate model.\footnote{We adopted the faint background model file pca\_bkgd\_cmfaintl7\_eMv20051128.mdl.}
We applied the GTIs to both observational data and simulated background
and extracted
lightcurves in initial 16s bins from channels that correspond to $\sim$2--20 keV.
The lightcurves of net count rates were calculated using \texttt{lcmath} in the HEASoft (v6.19) package.
Since only PCU2 among the five proportional counter units (PCUs) of PCA is almost always in operation,
we extracted lightcurves from PCU2 for flaring event selection.
We visually inspected every lightcurve to select events in individual observations that contain a complete or nearly complete sub-hour flaring profile.
Specifically, we require the flare to have apparent rise and fall; we also require the existence of a plateau either before the rise and/or after the fall
to assess the completeness of the flare and the level of the background component.\footnote{We do not adopt quantitative criteria to select
the events, automatically, because quantitative criteria are inescapable of subjective tweak and visual inspection is almost always necessary.}
Note that {\it RXTE} is in low Earth orbit,
and thus an uninterrupted lightcurve is usually less than $\sim50$ minutes due to the Earth's occultation or passage through the
South Atlantic Anomaly, etc., which
limits the timescales of the events investigated.
After identifying fast flaring events,
we extracted lightcurves and spectra from all the PCUs available during that observation
to achieve higher signal-to-noise ratio (S/N) for further analysis.

From the complete {\it RXTE/PCA} database, two new fast flaring events were found. They belong to \pks and \s5.
The lightcurves of these two events are shown in Fig.~\ref{fig:micro_flares}.
We also plot an event of \mrk{} \citep[reported in][]{Xue2005}.
A fast-flaring candidate event of \fes is reported in Appendix~\ref{sec:fes}, which has relatively low credibility because of limited S/N.
These flaring observations generally lack simultaneous observations in other wavebands.

We checked for potential contamination by soft electron flares that were not screened out by
the criteria ``ELECTRON2.LE.0.1'' in data cleaning.
The contemporaneous Electron2s of each event were well below 0.1
and did not show any apparent electron flaring activity that may be responsible for the X-ray flares.
The longitudes and latitudes of the satellite at the onsets of the X-ray flares did not
cluster in the anomalous high background region \citep[cf. Fig.~7 of][]{Xue2005}.
Further support for the genuineness of the flaring events comes from their lightcurve and spectral
features explored below. They behave like well-known X-ray flares of TeV blazars, only at much smaller timescales.
In conclusion, we did not find any sign of contamination of soft electrons or any other known sources for all the flaring events.
However, the possibility of an unrelated X-ray transient in the field of view still cannot be ruled out entirely since the PCA lacks the capacity of imaging.

\section{Lightcurve fitting}
\label{sec:lc}

\begin{table*}[!t]
    \caption{Fitting results for the fast flaring events}
\centering
\begin{tabular}{ccccccccccc} \hline \hline
    Band & Energy & $F_\textrm{c}$ & $F_0$ & $t_0$ & $\tau_\textrm{r}$ & $\tau_\textrm{d}$ & $t_\textrm{p}$ & $F_\textrm{p}/F_\textrm{c}$ & $\xi$ & $\chi_{\nu}^2 / dof$ \\
         &  (keV)& (c/s)& (c/s)& (s)& (s)&(s)&(s)& & \\
\hline
\multicolumn{11}{c}{\pks{}} \\
\hline
Soft & 1.94--5.47& $2.08_{-0.02}^{+0.02}$& $3.81_{-0.26}^{+0.28}$& $303_{-6}^{+8}$& $23_{-4}^{+5}$& $160_{-15}^{+16}$& $343_{-7}^{+8}$& 2.3 & 0.75 & 0.957/166 \\
Medium& 5.47--10.11& $0.98_{-0.02}^{+0.02}$& $3.45_{-0.24}^{+0.26}$& $318_{-7}^{+10}$& $29_{-4}^{+8}$& $141_{-15}^{+14}$& $356_{-6}^{+9}$& 3.2 & 0.66 & 1.05/166 \\
Hard & 10.11--20.30& $0.33_{-0.02}^{+0.02}$& $2.43_{-0.34}^{+0.35}$& $324_{-14}^{+33}$& $29_{-8}^{+18}$& $105_{-30}^{+25}$& $353_{-12}^{+19}$& 5.3 & 0.56 & 0.895/166 \\
Full & 1.94--20.30& $3.4_{-0.04}^{+0.04}$& $9.49_{-0.48}^{+0.47}$& $311_{-5}^{+6}$& $25_{-3}^{+4}$& $143_{-10}^{+11}$& $348_{-5}^{+6}$& 2.8 & 0.70 & 1.18/166 \\
\hline
\multicolumn{11}{c}{\s5{}} \\
\hline
Soft & 2.06--5.31 & $0.19_{-0.08}^{+0.05}$&$2.04_{-0.32}^{+0.28}$&$1344_{-100}^{+111}$& $222_{-43}^{+78}$&$825_{-171}^{+326}$&$1573_{-52}^{+66}$& 7.5 & 0.58 & 1.48/23\tablenotemark{(b)} \\
Hard & 5.31--10.11 & $0.00_{-0.08}^{+0.04}$&$1.67_{-0.37}^{+0.21}$&$1216_{-108}^{+64}$&$161_{-39}^{+65}$&$1588_{-285}^{+1903}$&$1551_{-54}^{+107}$&$>17$\tablenotemark{(a)} & 0.82 &1.62/23\tablenotemark{(b)}\\
    Full & 2.06--10.11 & $0.17_{-0.09}^{+0.08}$& $3.95_{-0.45}^{+0.38}$& $1258_{-55}^{+57}$& $178_{-29}^{+40}$& $1158_{-179}^{+320}$& $1547_{-39}^{+48}$& 17 & 0.73 &  1.15/159\tablenotemark{(b)} \\
\hline
\multicolumn{11}{c}{\mrk{} first substructure} \\
\hline
Soft & 1.94--5.82 & $5.60_{-0.04}^{+0.03}$ & $3.32_{-0.88}^{+0.02}$ & $1918_{-59}^{+32}$ & $99_{-34}^{+27}$& $110_{-41}^{+75}$ & $1924_{-26}^{+14}$ & 1.30 & 0.05 & 0.99/151 \\
Hard & 5.82--20.30 & $5.35_{-0.04}^{+0.05}$ &$3.29_{-0.58}^{+0.38}$ &$2001_{-80}^{+8}$ &$147_{-54}^{+18}$ &$33_{-6}^{+65}$ &$1961_{-40}^{+8}$ & 1.38 & $-0.64$ & 1.09/151\\
Full & 1.94--20.30& $10.96_{-0.06}^{+0.06}$& $6.9_{-1.1}^{+0.2}$& $1961_{-47}^{+18}$& $119_{-29}^{+21}$& $67_{-19}^{+41}$& $1936_{-18}^{+9}$ & 1.33 & $-0.28$ & 1.23/151 \\
\hline
\multicolumn{11}{c}{\mrk{} second substructure} \\
\hline
Soft & - & - & $1.78_{-0.58}^{+0.25}$ & $2206_{-23}^{+90}$ & $40_{-10}^{+135}$ & $160_{-65}^{+71}$ & - & - & 0.60 & - \\
Hard & - & - & $1.73_{-0.34}^{+0.35}$&$2135_{-16}^{+84}$&$30_{-11}^{+79}$&$302_{-115}^{+225}$& - & - & 0.82 & - \\
Full & - & - & $3.7_{-0.9}^{+0.6}$& $2177_{-18}^{+54}$& $43_{-13}^{+55}$& $213_{-63}^{+83}$&  - & - & 0.67 & - \\
\hline
\end{tabular}
\tablenotetext{1}{Note our fitting cannot constrain the hard-band quiescent flux level of \s5{}, so the hard-band amplitude is unmeasurable.
    We thus set the full-band amplitude as the lower limit of that of the hard band.}
\tablenotetext{2}{We fitted the model to the soft- and hard-band lightcurves in bins of 96s, while the full-band lightcurve is in bins of 16s.}
\label{tab:fit_result}
\end{table*}

\begin{figure}[!t]
\centering
\includegraphics[width=0.48\textwidth, clip]{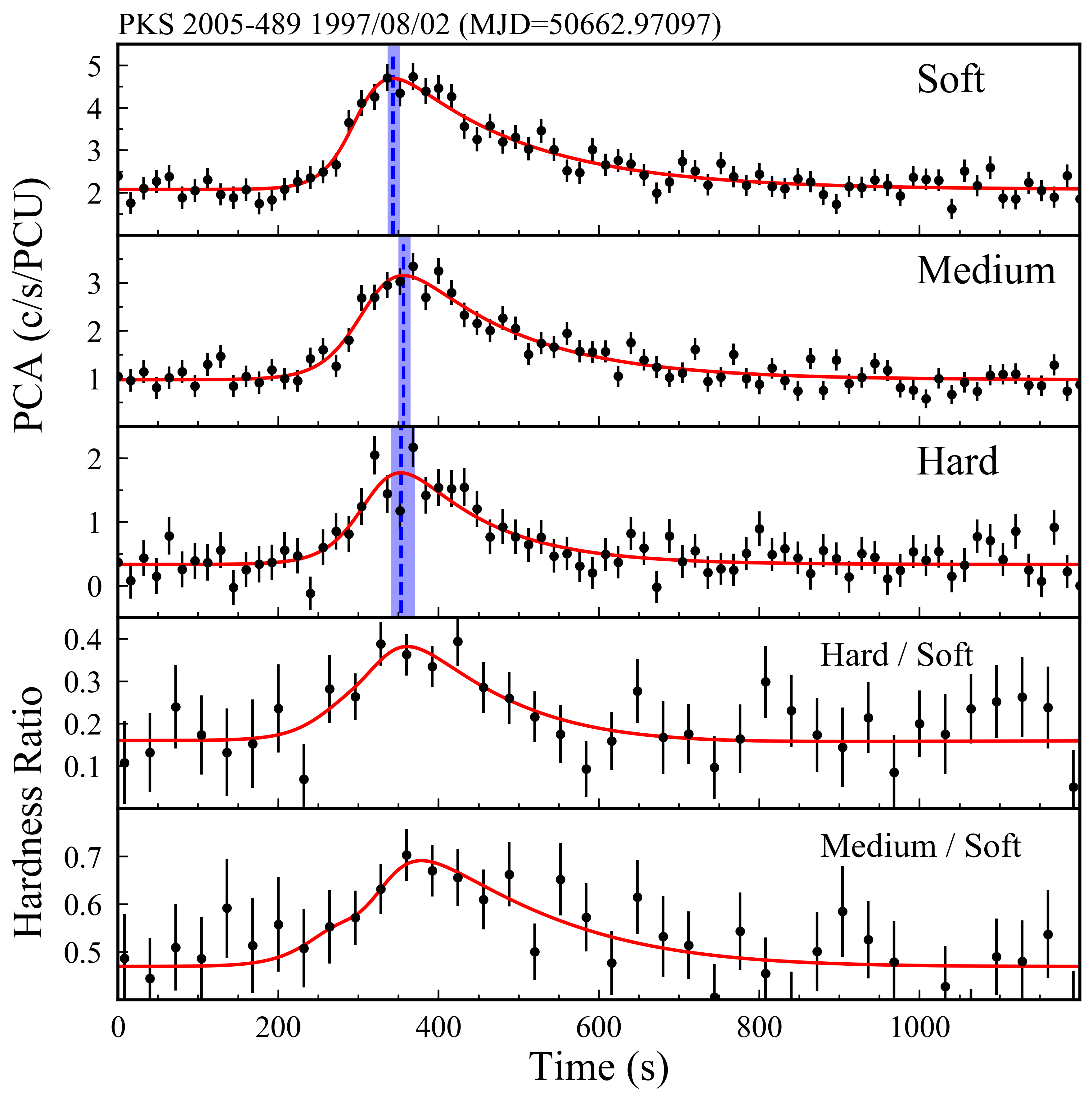}
    \caption{Top three panels: The rapid X-ray flare of \pks{} in three energy bands in bins of 16s.
    The red solid curves are weighted least square models;
    the vertical blue dashed lines indicate $t_\textrm{p}$,
    whose $1\sigma$ uncertainties are shown as the shaded blue regions.
    Bottom two panels: Hardness ratios (hard band to soft band and medium band to soft band);
    the red solid curves are hardness ratios calculated from the weighted least square models above. The data points are in bins of 32s.}
\label{fig:fit_pks}
\end{figure}

\begin{figure}[!t]
\centering
\includegraphics[width=0.45\textwidth, clip]{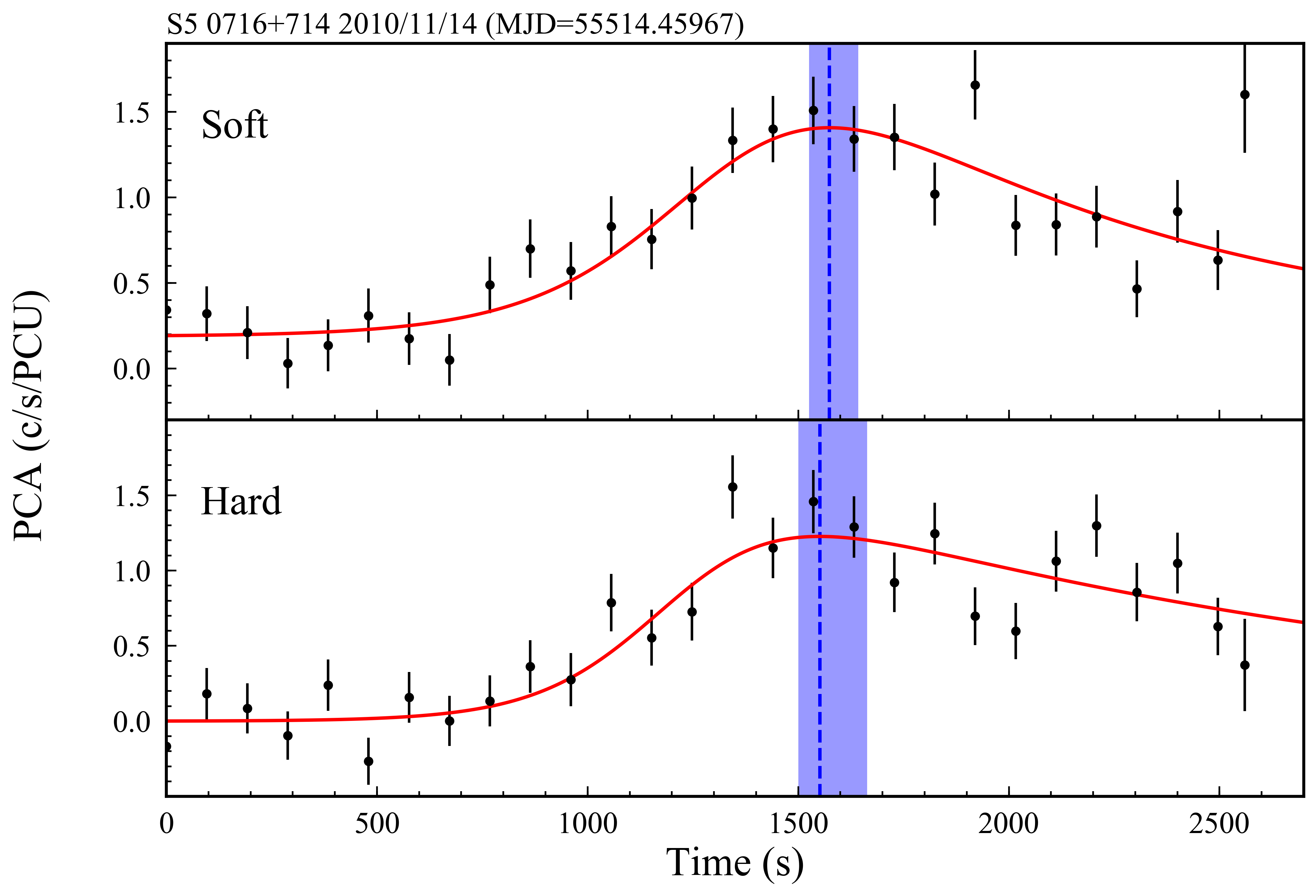}
\caption{Soft- and hard-band lightcurve fitting of \s5{}.
    The lightcurves are in bins of 96s.
The vertical dashed lines indicate $t_\textrm{p}$,
whose $1\sigma$ uncertainties are shown as the shaded blue regions.}
\label{fig:fit_s5}
\end{figure}

\begin{figure}[!t]
\centering
\includegraphics[width=0.45\textwidth, clip]{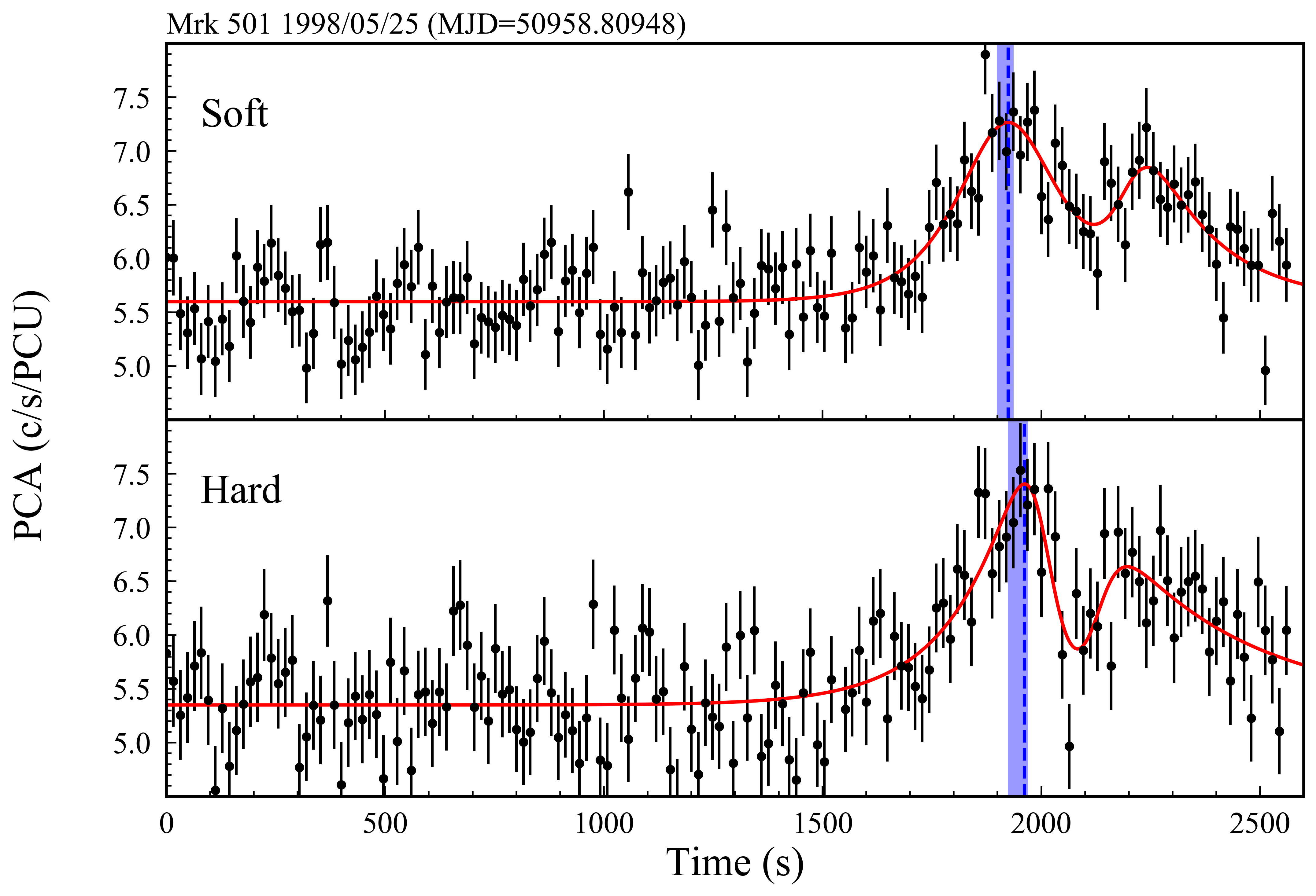}
    \caption{Soft- and hard-band lightcurve fitting of \mrk{}.
    The lightcurves are in bins of 16s.
    The vertical dashed lines indicate the peak of the main flare $t_\textrm{p}$,
    whose $1\sigma$ uncertainties are shown as the blue shaded regions.}
\label{fig:fit_mrk}
\end{figure}

Lightcurves in additional energy bands (see the second column of Table~\ref{tab:fit_result} and Figs.~\ref{fig:fit_pks},\ref{fig:fit_s5},\ref{fig:fit_mrk})
are extracted according to the energy-channel conversion
table.\footnote{See the table at http://heasarc.gsfc.nasa.gov/docs/xte/e-c\_table.html.}
Thanks to the high data quality,
variability at timescales down to the time resolution (16s) of the lightcurves (see Fig.~\ref{fig:fit_pks}) is seen.
Because of the variety of data quality and gain epochs \citep{Jahoda2006}, we do not have
uniform definitions for different bands for all three sources.
We have four different bands for \pks{}, while we have three bands for \mrk{} and \s5{};
the full band of \pks{} and \mrk{} is 1.94--20.30~keV, while the full band of \s5{} is 2.06--10.11~keV.

\subsection{Method of Fitting}

We fitted the lightcurves with a constant flux plus an exponentially rising and decaying flare
following \cite{Abdo2010b}\footnote{In addition to the frequently-used Eq.~\ref{eqn:flare_profile}, some similar analytical expressions have been used to describe the flare profiles of blazars;
see \cite{Albert2007, Giebels2007, Chatterjee2012}.}:
\begin{equation}
\label{eqn:flare_profile}
    F(t) = F_\textrm{c} + F_0\left(e^{\frac{t_0-t}{\tau_\textrm{r}}}+e^{\frac{t-t_0}{\tau_\textrm{d}}}\right)^{-1},
\end{equation}
where $F_\textrm{c}$ represents the constant flux level underlying the flare,
and $\tau_\textrm{r}$ and $\tau_\textrm{d}$
are the characteristic rising and decaying timescales\footnote{The doubling and halving timescales are $\tau_{\rm r}\times\ln2$ and $\tau_{\rm d}\times\ln2$.}
of the flare.
 $t_0$ indicates the transition from rising to falling, and the count rate actually peaks at
\begin{equation}
\label{eqn:tp}
    t_\textrm{p}=t_0+ \frac{\tau_\textrm{r}\tau_\textrm{d}}{\tau_\textrm{r} + \tau_\textrm{d}}\ln\left(\frac{\tau_\textrm{d}}{\tau_\textrm{r}}\right),
\end{equation}
which equals $t_0$ only when the flare is symmetrical ($\tau_\textrm{r}=\tau_\textrm{d}$). Therefore we define the amplitude of the flare as $F_\textrm{p}/F_\textrm{c}$,
i.e., the count rate at $t_\textrm{p}$ ($F_{\rm p}$) over the constant level, instead of $F_0/F_\textrm{c}$.
The symmetry of a flare is described by
\begin{equation}
\label{eqn:xi}
\xi=\frac{\tau_\textrm{d}-\tau_\textrm{r}}{\tau_\textrm{d}+\tau_\textrm{r}},
\end{equation}
whose value is in the range of [$-$1, 1].
$\xi=-1$ ($=1$) represents completely right (left) asymmetric profiles with a zero falling (rising) timescale; $\xi=0$ indicates a symmetric flare.

The uncertainties of $t_\textrm{p}$ and $F_\textrm{p}$ have to be propagated from the errors of other parameters.
We adopted an MCMC (Markov Chain Monte Carlo) algorithm to fit the lightcurves,
which returns reliable probability intervals of timescales and amplitudes by sampling from their posterior distributions.
We first performed weighted least squares fitting using a numerical minimizer to
obtain the best estimates of $F_\textrm{c}$, $F_0$, $t_0$, $\tau_\textrm{r}$,
and $\tau_\textrm{d}$. Starting from these initial values, we took 1000
random walk steps in parameter space.
The samples of $t_\textrm{p}$ and $F_\textrm{p}$ were calculated according to Eq.~\ref{eqn:tp} and Eq.~\ref{eqn:flare_profile}.
Note that we added a second flaring component to the model in the fitting of \mrk{} (see the bottom panel of Fig.~\ref{fig:micro_flares}).

\subsection{Correcting Lightcurve Error Bars}
\label{sec:corr_e}
The initial fits have reduced Chi-square values ($\chi_{\nu}^{2}=\chi^2/dof$) in the range of 0.48--0.80,
which indicates that the assigned error bars are larger than true statistical fluctuations.
Indeed, the standard \xte/PCA data reduction pipeline overestimates the lightcurve errors \citep{Nandra2000}.
The error estimation of the net lightcurves is propagated from the error estimation of the observed lightcurves and
the simulated background lightcurves, of which the latter is too smooth to be described by the assumed Poisson statistics.
We decided to correct the error estimation using $\sigma^2_{\rm net}=\sigma^2_{\rm obs}+k^2\sigma^2_{\rm bkg}$, where $0\le k^2 <1$.
The correction factor $k^2$ can be determined by forcing the excess variance\footnote{Excess variance is the variance after subtracting the mean square error \citep[e.g.,][]{Nandra1997, Vaughan2003}.}
of the quiescent parts in the top and bottom panels of Fig.~\ref{fig:micro_flares} to be zero.
The resulting correction factors of different segments at different energy bands span from $-0.13$ to 0.71.
We decided to fix $k^2=0$ as in \cite{Nandra2000}.
We have ignored errors on the background in the lightcurve analysis below, unless otherwise stated.
The fitting results after correcting the error bars are tabulated in Table~\ref{tab:fit_result}.
Note that we still report the least square results in Table~\ref{tab:fit_result} as the estimation of each parameter,
but the $1\sigma$ intervals are derived from MCMC fitting. The reduced Chi-square values of most fits are around 1.

\subsection{Lightcurve Fitting Results}

Every lightcurve shows some flare-like structure above a constant ``background'' flux level, which actually varies on longer timescales.
From the flux levels of the constant components, the events occur when the sources are in relatively high states,
but they do not always coincide with the periods with the highest flux levels.
For example, the \xte{}/PCA count rates of \pks{} can be 10 times the constant flux level here as found about one and a half years later \citep{Perlman1999}.

The variation amplitude is higher in harder bands,
which suggests that the flaring component has a harder spectrum than the corresponding constant component.
The mixed spectra become harder when the flux rises and the flaring component becomes more prominent.
We discuss the spectral variability of \pks{} in detail in Section~\ref{sec:pks_spec_v}.
We also fit the spectra of \pks{}, \s5{}, and \mrk{} and confirm that the flaring components have harder spectra in Section~\ref{sec:spec}.


There is a trend of rising timescales being shorter and decaying timescales being longer at softer energies.
This trend is obvious in \pks{} and the first flare substructure of \mrk{} (see Table~\ref{tab:fit_result};
see also Mrk 421 in Appendix~\ref{sec:mrk421}).
This suggests that the variability is caused by electron acceleration and cooling \citep[e.g.,][]{Fraija2017}.
As a consequence, the flares are more right asymmetric in harder bands (i.e., smaller $\xi$ values).
\s5{} and the second flare substructure of \mrk{} do not follow the patterns, although the error bars prevent any solid conclusions.
The difference of timescales between energy bands is not obvious for \s5{}, mainly because the decay of the flare was not completely sampled
and $\tau_\textrm{d}$ cannot be constrained well. Moreover, only one PCU was operating during the observation, so the S/N is low.

The vertical lines in Fig.~\ref{fig:fit_pks} and Fig.~\ref{fig:fit_mrk} suggest hard lags in the variability of \pks{} and \mrk{},
which means that the variation of hard photons lags that of soft photons.
The suggested lag of \pks{} is not confirmed by cross-correlation function (CCF), presumably due to the limited time-resolution (see discussion in Section~\ref{sec:pks_spec_v}).
Again, the soft-band and hard-band peaks of \s5{} do not show an obvious difference, putatively due to the larger error bars (see Fig.~\ref{fig:fit_s5}).

\subsection{Spectral variability of \pks{}}
\label{sec:pks_spec_v}

\begin{figure*}[!t]
\centering
\includegraphics[width=0.95\textwidth, clip]{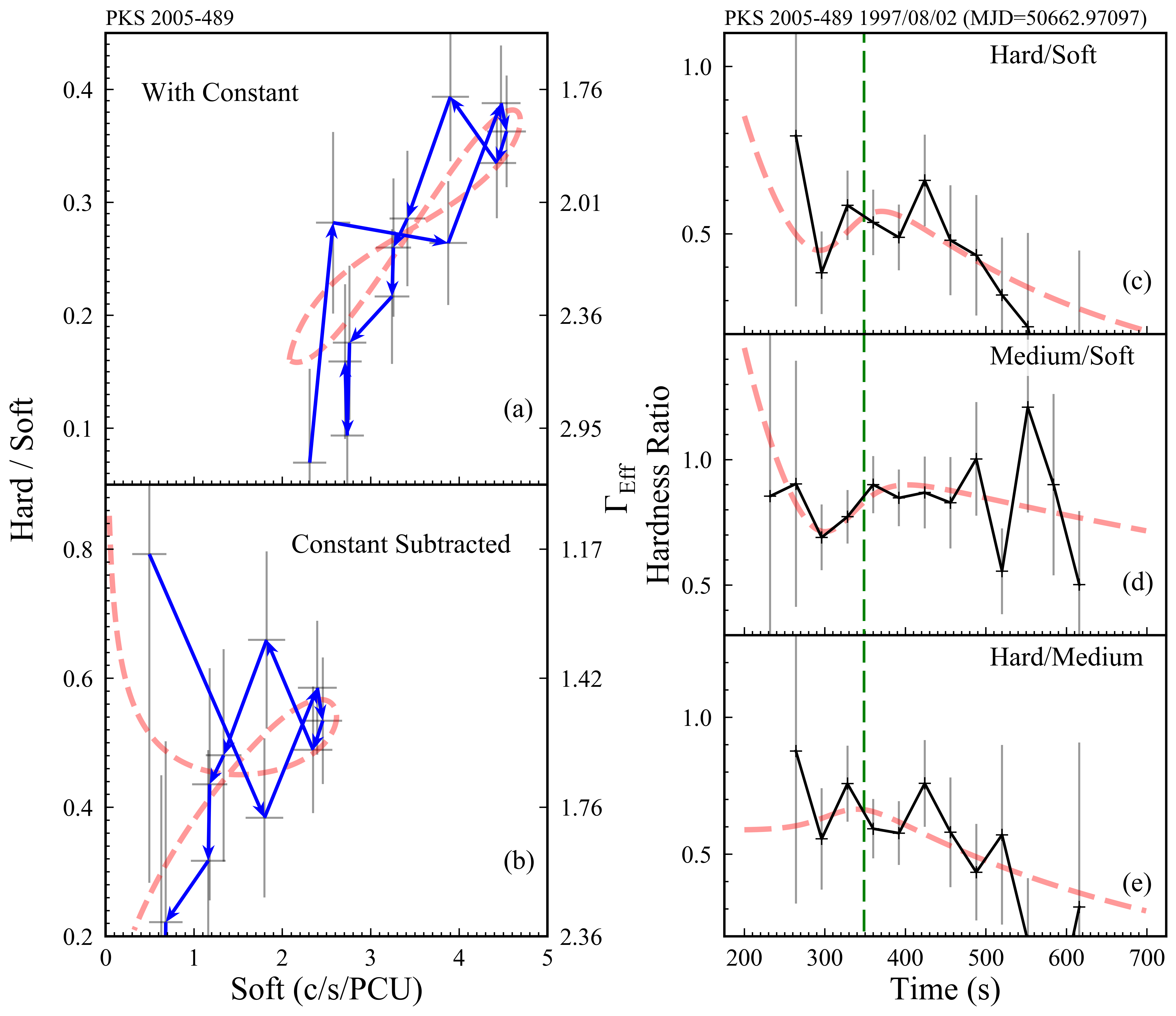}
\caption{Panels~(a)(b): HR--flux diagram of \pks{} before and after subtracting the constant component $F_\textrm{c}$. Blue arrows indicate the time sequence.
    The effective photon index $\Gamma$ shown as the $y$-axis on the right hand side is obtained using the response files of the PCA, a Galactic absorbed power-law model, and a range of assumed photon indices.
    Panels~(c)(d)(e): Three hardness ratios of the flare versus time after subtracting the constant component $F_\textrm{c}$; the vertical dashed lines indicate the time of full band peak ($t_{\rm p}$).
    We only considered the $\sim400$s (from $\sim200$s to $\sim600$s) segment which contains nearly the whole flare.
The data points are in bins of 32s.
    The dashed curves are calculated from analytical models (Eq.~\ref{eqn:flare_profile}).}
\label{fig:hr-flux_pks}
\end{figure*}

The flare of \pks{} has the shortest rising timescales and its data are of the highest quality,
so we investigated its spectral variability in detail.
We plot the hardness ratios\footnote{We define hardness ratio as the count rate of hard band over that of soft band, ${\rm\frac{H}{S}}$.
The errors are calculated as $\sigma_{\rm HR}={\rm \frac{H}{S}}\sqrt{\big(\frac{\sigma_{\rm H}}{{\rm H}}\big)^2+\big(\frac{\sigma_{\rm S}}{\rm S}\big)^2}$.}
(HRs) in the bottom two panels of Fig.~\ref{fig:fit_pks}, which show hardening that corresponds to the flare.
We also plot a HR--flux diagram of this event
in panel~(a) of Fig.~\ref{fig:hr-flux_pks},
which shows a ``harder when brighter'' trend and hysteresis.
The loop begins with clockwise motion, and then follows a counterclockwise direction.
Below, we argue that the clockwise trend at the beginning is due to the superposition of two spectral components.

The apparent two-component nature of the lightcurves suggests that the spectral variability is
partially caused by a change of the relative fraction of the two components.
In other words, if the flaring component has different hardness ratio from that of the constant component,
then even if neither of the spectra changes over time, the observed overall hardness ratio will still change
due to the flux variation of the flaring component \citep[see][]{Sun2014, Ramolla2015}.
However, the hysteresis loop in panel~(a) suggests the spectrum of the flaring component is intrinsically variable;
otherwise the track in the HR--flux plane while the flux is rising will be identical to the track while the flux is declining, instead of forming a loop.
To discriminate the
effects caused by the mixing of different components and the flare's intrinsic spectral variability,
we further subtract the constant component $F_\textrm{c}$ from the
lightcurve of each band, where $F_\textrm{c}$ is from Table~\ref{tab:fit_result}.
We calculated the hardness ratios from the resulting flare-only lightcurves in panels~(c)(d)(e)
and HR--flux diagram in panel~(b).
The plots suggest that the
flare emerges with a hard spectrum, softens gradually as the flux rises, then hardens near
the highest flux, and finally softens as the flare fades away.
Only a counterclockwise loop is apparent in the HR--flux plane for the flare-only lightcurve.
Based on the evidence above, we know that the clockwise loop in panel~(a) could be caused by the sudden emergence of a hard flaring component
and the spectral variability soon
follows the spectral variability of this flaring component due to its increasing dominance.
The flaring component itself has complex spectral variability and the counterclockwise
loop in panel~(b) suggests the existence of a hard time lag, albeit being subject to large uncertainties.
Note that the overall oblique ``8'' shape before subtracting the constant component in panel~(a)
is reminiscent of a flare of Mrk~421 which lasted $\sim60$ ks and was reported by \cite{Garson2010} using {\it Suzaku} data.
We reanalyzed the {\it Suzaku} data of Mrk~421 and
performed the same spectral variability analysis as for \pks{} above in Appendix~\ref{sec:mrk421}.
We reproduced the same results as for \pks{} in the flare of Mrk~421 and thereby strengthened our conclusions.
The similar behavior of spectral variability, seen both in the extremely rapid flare of PKS 2005-489 and
in the long-duration flare of Mrk 421,
is reminiscent of the scale invariant nature of X-ray flares from TeV blazars \citep{Cui2004, Xue2005}.

We also calculated the CCF \citep{Edelson1988, Welsh1999}
of the soft band and hard band flare-only lightcurves, but did not find an obvious time lag.
Any time lag in this fast flaring event of \pks{} could be intrinsically small compared with
the time resolution of the observation (16s).
The bandpass of our data is narrow, spanning about one order of magnitude, so any energy-dependent lags may
not be significant.
The size of lags may also be positively correlated to the duration of the flares \citep{Zhang2002}, which
means that flares of shorter duration have smaller lags.
Note that in Table~\ref{tab:fit_result} the differences of $t_0$, $t_\textrm{p}$, and $\tau_\textrm{r}$ between different bands are
small compared with relatively large error bars.

\section{Spectral Fitting}
\label{sec:spec}

\begin{figure*}[!t]
\centering
\includegraphics[width=0.95\textwidth, clip]{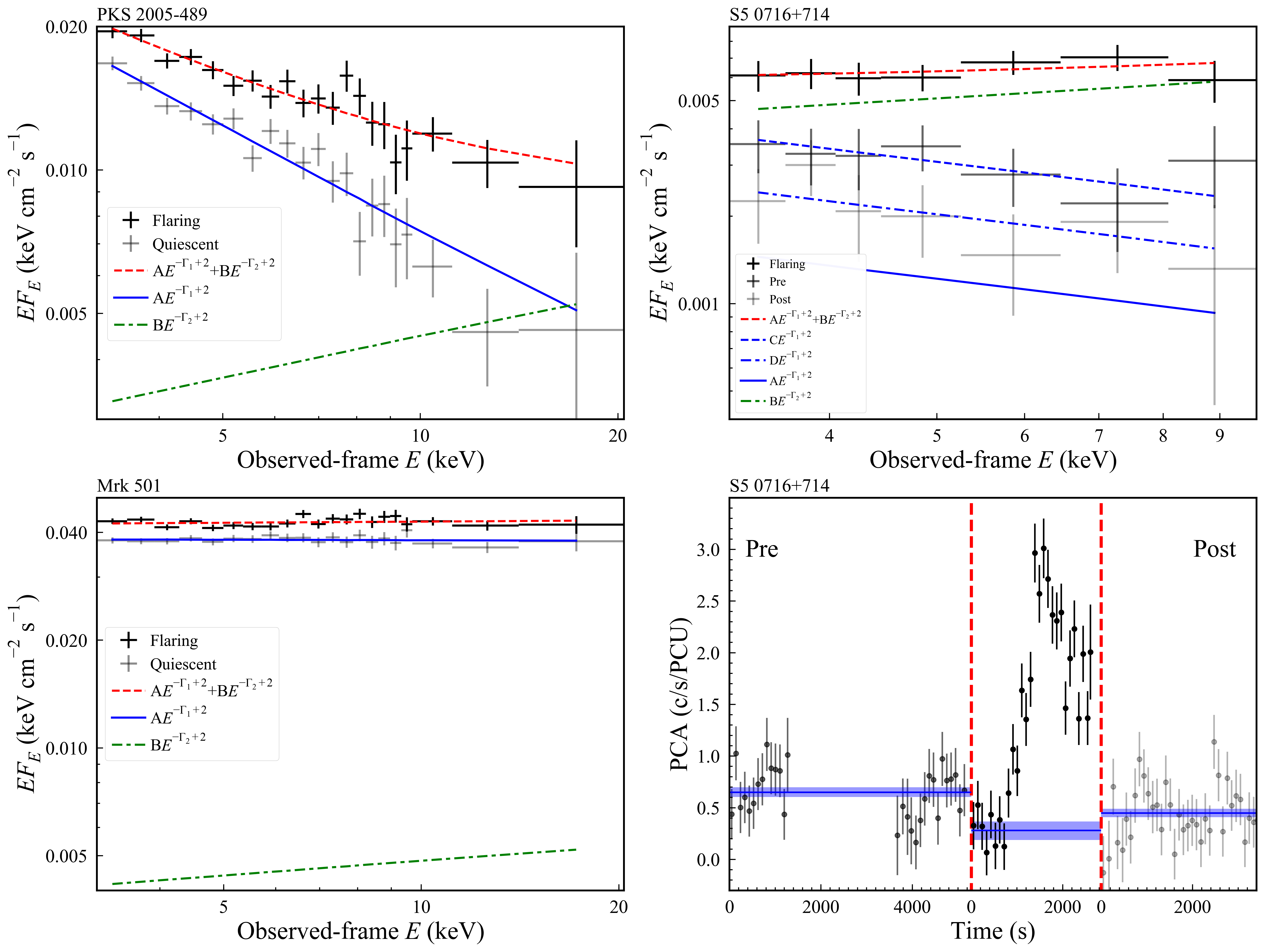}
\caption{In the spectral fittings (the top row and bottom-left panel),
the Galactic absorption is fixed to $N_\textrm{H}=5.08\times10^{20}\;\textrm{cm}^{-2}$,
$3.81\times10^{20}\;\textrm{cm}^{-2}$ for \pks{}, \s5{} \citep{Dickey1990} and $1.56\times10^{20}\;\textrm{cm}^{-2}$ for \mrk{} \citep{Kalberla2005}.
Note that the average spectra of the flaring components are shown in green dash-dotted lines.
In the bottom-right panel, we show the observations that were taken before the flare (pre) and after the flare (post) of \s5{},
whose ObsIDs are 95377-01-90-00 and 95377-01-92-00. Note that the vertical dashed lines represent gaps of 3--4 days between adjacent observations.
We fit the pre-flare and post-flare lightcurves with constant fluxes
and show the results by horizontal lines. Shaded regions indicate $1\sigma$ uncertainties.}
\label{fig:spectrum}
\end{figure*}

\begin{table*}[!t]
    \caption{Spectral fitting results of the fast flaring events}
\centering
\begin{tabular}{ccccccccc} \hline \hline
    Source &  $\Gamma_{1}$ & $\Gamma_{2}$ & $F_{\textrm{quiescent}}$\tablenotemark{(a)} & $F_\textrm{flare}$\tablenotemark{(a)} & $F_{\textrm{flare only}}$\tablenotemark{(a)} & $\chi^2_{\nu}/dof$ & $L_\textrm{X}$\tablenotemark{(b)} & $E$\tablenotemark{(c)} \\
\hline
\pks{} & $2.73\pm0.05$ & $1.71\pm0.18$ & 29.5& 42.4& 12.9& 0.88 / 38 &  $3.86\times10^{44}$& $2.08\times10^{47}$\\
\hline
\mrk{} & $2.01\pm0.02$& $1.86\pm0.20$& 115.7 & 130.0 & 14.3 & 0.76 / 38 &  $3.23\times10^{44}$& $5.13\times10^{46}$ \\
\hline
\s5{} &  $2.48\pm0.27$ & $1.89\pm0.15$& 5.7 (pre) & 12.3 & 12.2 & 0.32 / 16 &  $1.90\times10^{45}$ & $1.06\times10^{49}$ \\
      & & & 3.8 (post) & & & & $1.25\times10^{45}$ & \\
\hline
\end{tabular}
    \tablenotetext{1}{Fluxes are in units of $10^{-12}$ ergs cm$^{-2}$ s$^{-1}$;
    the energy range is 3--20 keV for \pks{} and \mrk, but 3--10 keV for \s5{}.}
    \tablenotetext{2}{X-ray luminosity of the constant component luminosity in units of ergs s$^{-1}$.}
    \tablenotetext{3}{Energy of the flaring component in units of ergs.}
\label{tab:spec_fit}
\end{table*}

Motivated by the two-component model in lightcurve fitting,
we carried out spectral fitting in a similar way.
By dividing the observation into flaring and quiescent phases according to the lightcurves,
we separately extracted and jointly fitted the spectra of the two phases.
The spectra were extracted from the top layers of the operating PCUs.
The spectra and models are shown in Fig.~\ref{fig:spectrum}
in the $EF_E$ representation and the photon indexes are tabulated in Table~\ref{tab:spec_fit}.
We used channels that correspond to 3--20 keV for \pks{} and \mrk{} and 3--10 keV for \s5{}, respectively.
In addition to Galactic absorption, we used a power law and sum of two power laws in {\sc XSPEC (v12.9.0)}
to fit the spectra of the quiescent and flaring phases, respectively.
We tied the power law in the quiescent phase to one of two power laws in the flaring phase,
for both normalization and photon index (see the legends of Fig.~\ref{fig:spectrum}).
We also calculated the X-ray luminosity of the constant component and the total energy of the flaring
component in the X-ray band (namely average flare luminosity times the length of the defined flare phase) in Table~\ref{tab:spec_fit}.
Note that the spectra should suffer from overestimation of error bars due to the same reason discussed in Section~\ref{sec:corr_e},
but we are not able to correct the error estimation of the spectra as we did in the lightcurve fitting.
One consequence of the overestimation of error bars is that the parameters in the spectral fitting have large confidence intervals.
Another consequence is that we might be able to fit the spectra with many models.
For example, we can fit the spectra of the flaring phase, quiescent phase, and whole observation with simple power laws,
and the Chi-square statistics can still be acceptable.
We think that describing the spectra as a combination of different power laws is more physically appropriate and
is consistent with the lightcurve fitting.

Note that due to the limited length of the observation, we have neither a complete flaring nor a quiescent phase of \s5{}.
Therefore we extracted the spectra of the two observations just before and after the flare,
which are referred to as pre-flare and post-flare, respectively.
The gaps between the observations are 3--4 days.
The three observations are plotted in the bottom-right panel of Fig.~\ref{fig:spectrum},
where we fitted each of the pre-flare and post-flare lightcurves with a constant flux.
The spectra are jointly fitted with a tied ``constant'' component photon index,
but the normalizations of the ``constant'' component are free parameters in this case.
In other words, the constant component varies on timescales of a few days.
If we leave the normalization of the constant component underlying the flare (factor ``A'' in the legend of top-right panel) free,
the value of ``A'' in the fitting results is negligible but its error bars are large.
Additionally, we are not able to constrain the photon index of the flaring component ($\Gamma_2$) very well in this case.
Thus we fixed $\textrm{A}=0.6\textrm{D}$, according to the lightcurve fitting results in the bottom-right panel of Fig.~\ref{fig:spectrum}.
The choice of 0.6 or another reasonable value does not affect the result of $\Gamma_2<\Gamma_1$.

The above fitting shows that for all three flares, the flaring components have harder photon indexes than the corresponding constant components (see Table~\ref{tab:spec_fit}).
The difference is not as apparent for \mrk{} because of the relatively small variation amplitude (i.e., relatively weak flaring component).
The photon index of the constant component ranges from 2.0 to 2.7,
indicating that we are observing different declining parts of the synchrotron hump for different sources.
The photon indexes of the flaring components, on the other hand, lie in a narrower range around 1.8.
Every estimated average flux (averaged over the length of the flare phase we defined) of the events
is $\sim10^{-11}$ ergs cm$^{-2}$ s$^{-1}$ (see Table~\ref{tab:spec_fit}).
Using the ``efficiency limit'' of compact sources \citep[e.g.,][]{Fabian1979, Brandt1999}, $\Delta L/\Delta t\lesssim2\times10^{41}\eta_{0.1}$ erg s$^{-1}$,
the matter to energy conversion efficiency $\eta$ is greater than 1 for \pks{} and \s5{}, suggesting the existence of significant boosting of the emission \citep[e.g.,][]{Remillard1991}.
Due to the small flare amplitude, the efficiency of \mrk{} does not exceed 1.

\section{discussion}
\label{sec:discussion}

\subsection{The Flaring and Constant Components}
We interpret all the lightcurves that we analyzed in detail (including Mrk~421 in Appendix~\ref{sec:mrk421}) as a superposition of a
constant component and a flaring component \citep[e.g.,][]{Fraija2017}.
The two-component model depicts a simpler scenario than a single component undergoing an outburst.
For the case of a single emitting region, the variation in the HR--flux plane is more complex.
Furthermore, there must have been a sudden enhancement of some key physical quantity
and this quantity has to fall back later to its value preceding the burst.

\subsection{X-ray Radiation Process of S5~0716+714}
S5~0716+714 is an IBL \citep{Ackermann2011} with a synchrotron hump peaking at optical wavelengths \citep{Anderhub2009}.
It has frequent intra-day variability and hysteresis loops in the color--magnitude plane in the optical band \citep[e.g.,][]{Man2016}.
\cite{Pryal2015} reported two rapid flares of S5~0716+714 in the X-ray band with low significance.
The X-ray emission covered by the \xte bandpass could be a mixture of both synchrotron and inverse-Compton emission \citep{Wierzcholska2016a}.
As such, S5~0716+714 will not necessarily behave the same way as HBLs in X-rays.
Depending on the state (low or high) of the source when the observation was made,
the dominant X-ray radiation process may change accordingly.
The spectrum may steepen when \s5{} brightens
since the synchrotron emission in the soft X-rays becomes increasingly important \citep{Giommi1999}.
However, the synchrotron component itself hardens due to the shifting of the synchrotron peak \citep{Ferrero2006, Zhang2010}.
The steep photon index and the high flux levels (3--10 keV) of pre-flare and post-flare observations are consistent with synchrotron emission.
The flaring component has a slightly inverted spectrum,
but the even higher flux level and short timescales still support an origin as synchrotron emission.
Therefore the flare we observed from S5~0716+714 does not seem to have a different radiation process from HBLs.

\subsection{Constraints on the Emission Region}
\label{sec:constraints}
\begin{figure}[!t]
\centering
\includegraphics[width=0.48\textwidth, clip]{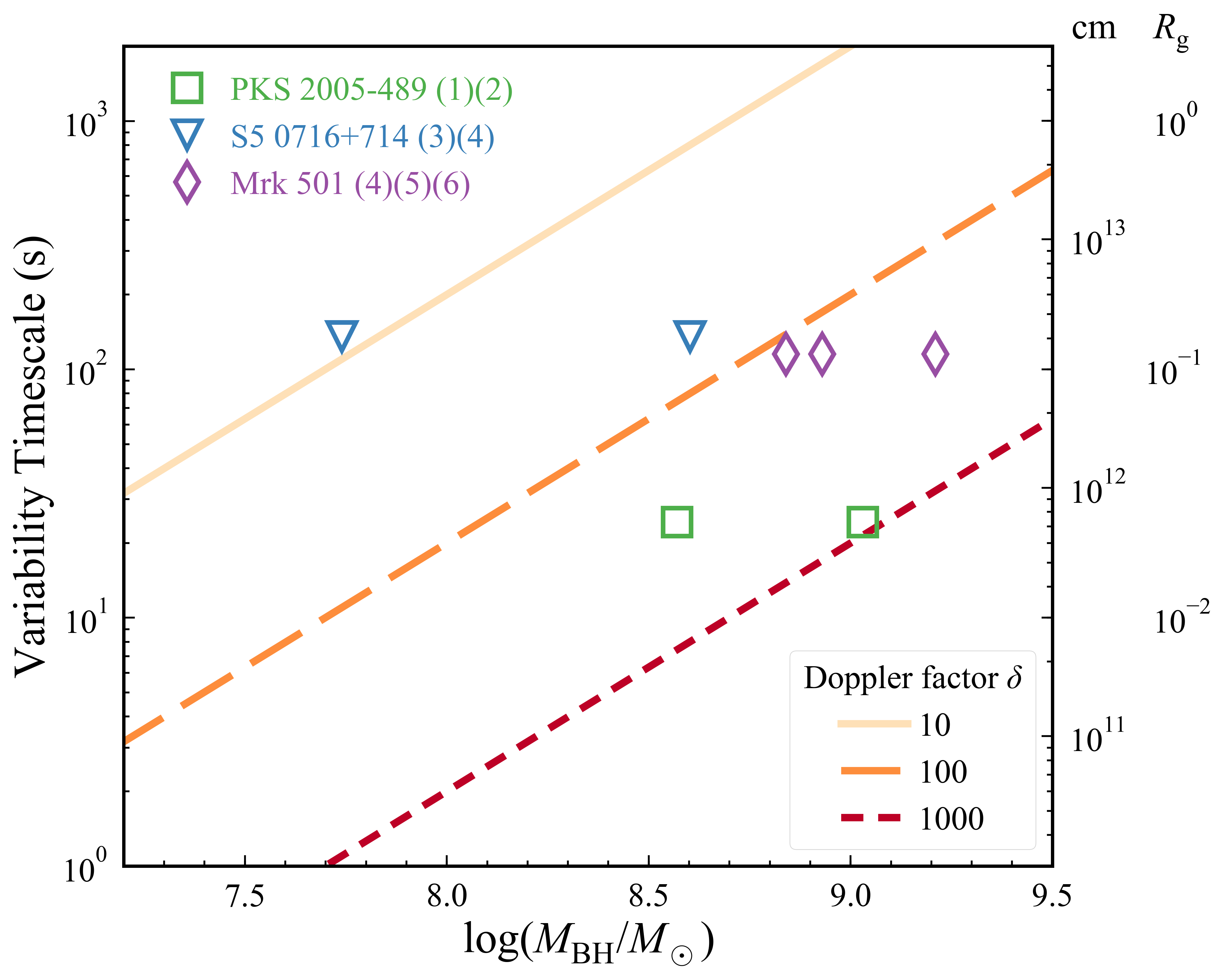}
    \caption{Variability timescale versus supermassive black hole mass.
    We use the rising timescales in the galaxy's frame, $\tau_{\rm r}/(1+z)$, as the values on the $y$-axis.
    For convenience, we annotate on the right $y$-axis the corresponding sizes, $\tau_{\rm r}c/(1+z)$, in units of cm and gravitational radius for $M_{\rm BH}=10^8M_\odot$.
    We also calculate light-crossing times for Kerr black holes, $t_\textrm{lc}=2 {\rm G} M_{\rm BH}/c^3 = 2\times10^3 \big(M_{\rm BH}/10^8M_\odot\big) {\rm s}$.
    The lines represent light-crossing times after considering the relativistic Doppler effect, $t_{\rm lc}/\delta$.
References of $M_{\rm BH}$: (1) \cite{Wagner2008}, (2) \cite{Woo2002}, (3) \cite{Wu2009}, (4) \cite{Ghisellini2010}, (5) \cite{Falomo2002}, (6) \cite{Barth2003}.}
\label{fig:mass_time}
\end{figure}

The upper limit on the physical scale of the flaring region can be given by:
\begin{equation}
    R \approx \frac{ct_{\rm flare}\delta}{(1+z)} \approx 10^{14}\left(\frac{\delta}{30}\right)\left(\frac{t_{\rm flare}}{100\;{\rm s}}\right) {\rm cm},
\end{equation}
where $\delta$ is the Doppler boosting factor, and $t_{\rm flare}$ is the observed variability timescale.
Since the flares are asymmetric, the size of the emitting region is reflected by the rising timescale $\tau_\textrm{r}$ \citep{Zhang2002}.
Searching the literature, we find black-hole mass estimates for three out of the four sources (except \fes{}).
We plot the rising timescale in the galaxy's frame against black-hole mass in Fig.~\ref{fig:mass_time}.
Note that the typical Doppler factor of BL Lac objects is 10--20 (20--30 for FSRQs) \citep[e.g.][]{Hovatta2009}.
The size of the black hole is often thought to be a natural lower limit on the physical scale of the emission region.
We also plot this lower limit as a function of black-hole mass assuming different Doppler factors.
The variability timescales should lie above the lines, if the size of the black hole is a hard lower limit.
The rising timescale of \pks{} is, as far as we know, shorter than any variability timescale
of AGNs at any wavelength ever reported \citep[e.g.,][]{Remillard1991,Yaqoob1997, Gallo2004, Xue2005,Aharonian2007,Albert2007,Aleksic2014,Kara2016}.
Fig.~\ref{fig:mass_time} shows that the Doppler factor of \pks{} has to be larger than several hundred, which appears unrealistic,
in order to support the idea that the black hole sets a lower limit on the physical size of the flaring region.
So far, there is no correlation found between the observed minimum
variability timescale and the black-hole mass \citep{Wagner2008, Vovk2015}.
The events of \pks{} and Mrk~421 (Appendix~\ref{sec:mrk421}) display remarkably similar spectral evolution,
supporting the same process driving the spectral variability.
However the timescales of the two events are different by two orders of magnitude,
reflecting an intrinsic difference between the two jets.
Noticeably, the famous Mrk~421 has many more X-ray observations than \pks{};
however, the variability timescale of Mrk~421 has never been found to be as short as the \pks{} event \citep{Cui2004, Pryal2015, Paliya2015}.
A lower limit on the variability timescale probably does exist, and probably is not set by the central supermassive black hole.

The synchrotron cooling time of emitting electrons is given by $t_{\rm cool}\approx6\pi m_{e}c/\sigma_{\rm T} \gamma B^2$ \citep{Rybicki1979}.
The observed photon energy at the synchrotron peak is given by $E_{\rm p}=\delta h \nu \equiv (3eh/4\pi m_e c)\delta \gamma^2 B$ \citep{Rybicki1979}.
Combining the above two equations, we have
\begin{equation}
t_{\rm cool} = 3.04\times10^3B^{-3/2}\delta^{-1/2}E_{\rm p}^{-1/2}\;{\rm s},
\end{equation}
where $E_{\rm p}$ is in units of observed keV \citep{Zhang2002}. If we take \pks{} as an example, and adopt $t_{\rm cool}=\tau_{\rm d}\sim143$~s, $E_{\rm p}=10$ keV and $\delta=30$,
we can have $B\approx1.2$~G. The estimated magnetic field of \mrk{} is similar to that of \pks{}, but the magnetic field of \s5{} is 4 times weaker.
Radio-loud AGNs are potential accelerators of cosmic rays.
We estimated the maximum energy of protons, if they can be accelerated in the same region as the electrons,
as $E_{\max}=eBR\sim$(2--9)$\times10^{16}$ eV.

\subsection{The Rarity of Extremely Rapid Flares of Blazars}
It is surprising to find such rapid variability of \pks{},
because it was not found to be variable on timescales less than a day \citep{Perlman1999, Rector2003, hess2010, hess2011}.
Out of the $\sim$160 PCA pointings on \pks{},
only two have positive excess variance\footnote{We do not correct the error bars of the
light curves here, but the light curves are consistent with being flat under visual inspection.}, including the one shown in Fig.~\ref{fig:micro_flares}(a).

Indeed, although the list of blazars that show extremely rapid flares is growing,
these events are rare \citep[e.g.,][]{Feigelson1986} and usually unexpected.
The rate of occurrence of sub-hour flares is about once per 4 Ms in the \xte{}/PCA database.
Most of the sources have only one such event reported, either in $\gamma$-ray or X-ray (e.g., \citealt{Gaidos1996, Aharonian2007}; and this paper).
These rapid flares do not seem to be the extremely short cases from a continuous distribution of the timescale of flares \citep[e.g.,][]{Li2017, Sasada2017}.
It remains unknown whether such events exist in all wavebands,
so we do not know the total energy output of the flares.
The biggest challenge is to coordinate multiple instruments to target the
same source simultaneously and hope rare unpredictable flaring events happen.

\subsection{Particle Acceleration}
Assuming the fast-cooling regime, the photon index of the flaring component ($\Gamma=$1.7--1.9) indicates that
the accelerated electrons have an effective energy spectral index ($p=2\Gamma-2$) in the range of 1.4--1.8,
which can be achieved by relativistic magnetic reconnection \citep[e.g.,][]{Guo2014}.
Indeed, the SED of \pks{} shows very low Compton dominance \citep{hess2010,hess2011}, which
indicates high magnetization. The results of SED modeling \citep[e.g.,][]{Anderhub2009, Abdo2011a, hess2011, Aleksic2015a}, which
usually assume one single homogeneous emitting region,
have low-strength magnetic fields ($B\sim$ 0.01--0.1 G) compared with the estimation of Section~\ref{sec:constraints} and below equipartition.
This may require that either the jet is structured \citep{Ghisellini2005} or only the regions that are responsible for
the fast flares we observed have high magnetization.

The direction of the loops in the HR--flux plane is thought to be determined by the competition between the
acceleration/ejection timescale, cooling timescale, and escape timescale \citep{Kirk1998}.
In practice, it is actually determined by which part of the synchrotron spectrum we are observing.
The distinctive pattern in the spectral variation of the flares of \pks{} and Mrk~421 is not predicted by
time-dependent homogeneous one-zone models \citep[e.g.,][]{Kirk1998, Chiaberge1999}.
The spectral variability pattern in Figs.~\ref{fig:hr-flux_pks} and \ref{fig:mrk421hr}
could possibly be produced by time-dependent inhomogeneous blazar models \citep[e.g.,][]{Bottcher2010}.

\section{Summary}
\label{sec:summary}

We searched the entire \xte{} archival database for rapid \mbox{X-ray} flares of TeV blazars that last less than one hour.
We investigated the temporal and spectral properties of the fast flares discovered under a two-component assumption.
Our analysis also includes an X-ray flare of Mrk~421 using {\it Suzaku} data
that has similar spectral variability to that of \pks{} in Appendix~\ref{sec:mrk421}.
Our main findings are as follows:

1. We discovered two new fast X-ray flares from \pks{} and \s5{} and a candidate flare from \fes{}.
The event of PKS~2005$-$489 shows, as far as we know, the most rapid variation of AGNs that has been observed at any wavelength.
The extremely small timescale ($\tau_{\rm r}<30$ s) defies the size that corresponds to the light-crossing time
of the supermassive black hole as a lower limit on the size of the flaring region.

2. The flares are usually superimposed on a constant/slowly-varying component.
The flaring component generally has a harder X-ray spectrum ($\Gamma=$ 1.7--1.9) than the constant component ($\Gamma=$ 2.0--2.7).

3. The higher data quality of the X-ray observations can provide more detail than $\gamma$-ray observations.
The flare-only component shows a counterclockwise pattern in the HR--flux diagram, providing a sign of likely hard lags.
This component also has the hardest spectrum right at its appearance. \\\\

\xte{} ceased science operation in January 2012, but X-ray observatories like {\it Chandra}, {\it XMM-Newton}, {\it Swift}, and {\it NuSTAR} are still accumulating
exposures on TeV blazars. For example, {\it XMM-Newton} has 9 Ms of exposure on TeV balazars as of July 2017.
A natural follow-up work is to search for additional rapid X-ray flares from TeV blazars in these archival databases.
With the increased sample, we may study more details of the spectral variability of the flares, thus shedding light on their origin and particle acceleration processes.
We can also study the correlations between flares and the properties of the jets and central engine, as well as the reason for the small flaring timescales.

We thank the anonymous referee for his/her helpful comments that improved the paper.
S.F.Z., Y.Q.X., and Y.J.W. acknowledge support from the 973 Program
(2015CB857004), NSFC-11473026, NSFC-11421303,
the CAS Frontier Science Key Research Program (QYZDJ-SSW-SLH006),
and the Fundamental Research Funds for the Central Universities.
S.F.Z. and W.N.B. acknowledge support from Chandra X-ray Center grant G04-15093X.

\appendix

\section{A. A candidate Fast flare of \fes{}}
\label{sec:fes}

\begin{figure}[!t]
\figurenum{A}
\includegraphics[width=0.45\textwidth]{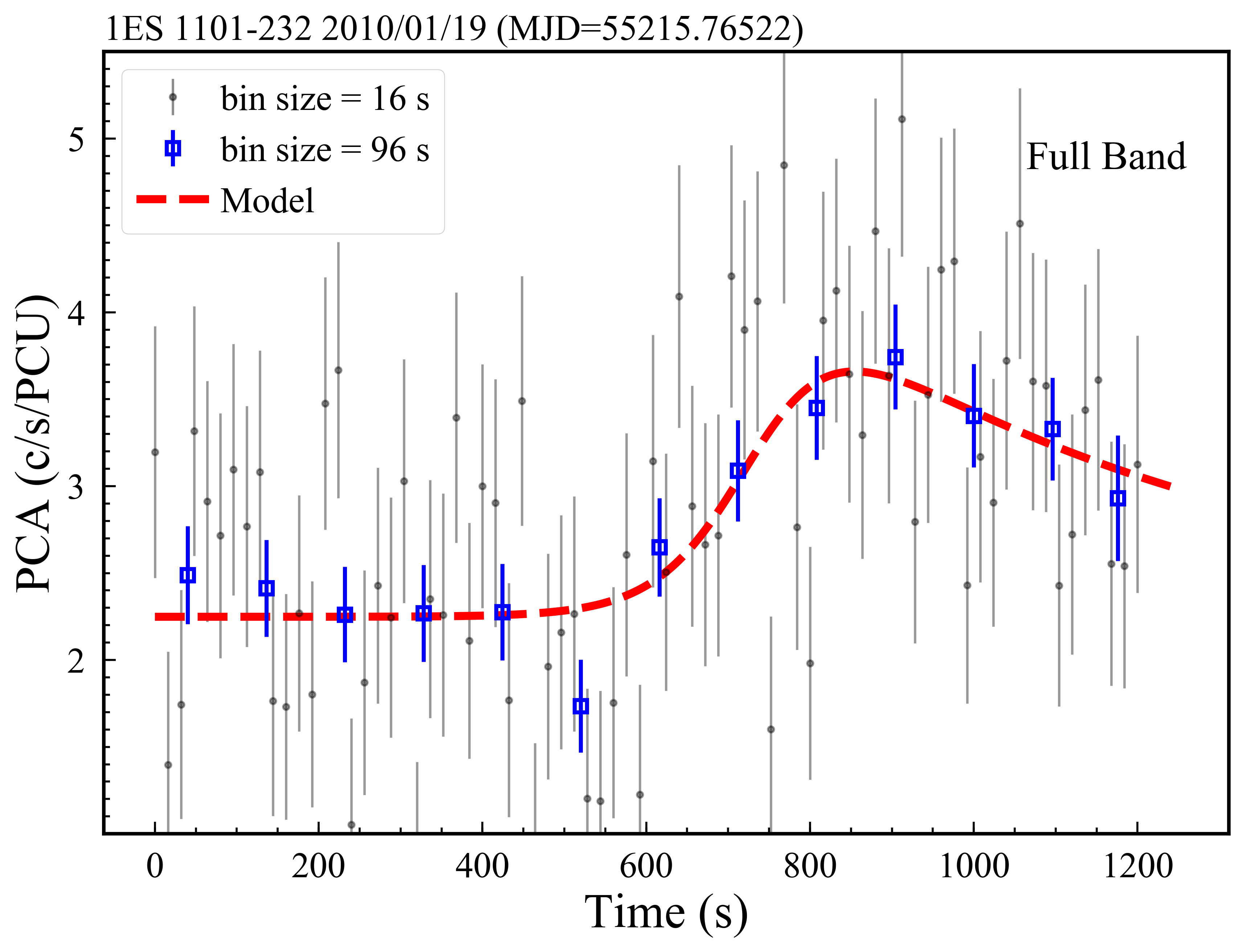}
    \caption{The candidate rapid X-ray flare of \fes{}.
    The ObsID is 95387-02-07-00, and only PCU2 was in operation during the observation.
The grey points and blue squares are the data of 16 s bins and 96 s bins, respectively.
The red dashed curve is the least square model fitting to the 96~s-bin light curve.}
\label{fig:1es_can}
\end{figure}

We found one fast X-ray flaring event of \fes{}, which is an HBL at a relatively high redshift ($z=0.186$) compared with
other HBLs.
The full-band (2--20 keV) lightcurve in 16 s bins (Fig.~\ref{fig:1es_can}) shows elevated flux in the second half of the observation.
When shown in 96 s bins, the lightcurve clearly manifests an almost complete flare profile,
and can be fitted using the model of Eq.~\ref{eqn:flare_profile}.
The amplitude ($F_{\rm p}/F_{\rm c}=1.93$) of this flare is modest, and the rising timescale ($\tau_{\rm r}=60$s) is almost as fast as \pks{}.
We only reported it as a candidate because the flux preceding the flare seems to be in continuous declining.
The noisy data prevent us from performing data analysis in the same fashion as other sources (sub-band lightcurve fitting, joint spectral fitting, etc.).

\section{B. Spectral variability of Mrk 421}
\label{sec:mrk421}
\begin{figure}[!t]
\figurenum{B1}
\includegraphics[width=0.40\textwidth]{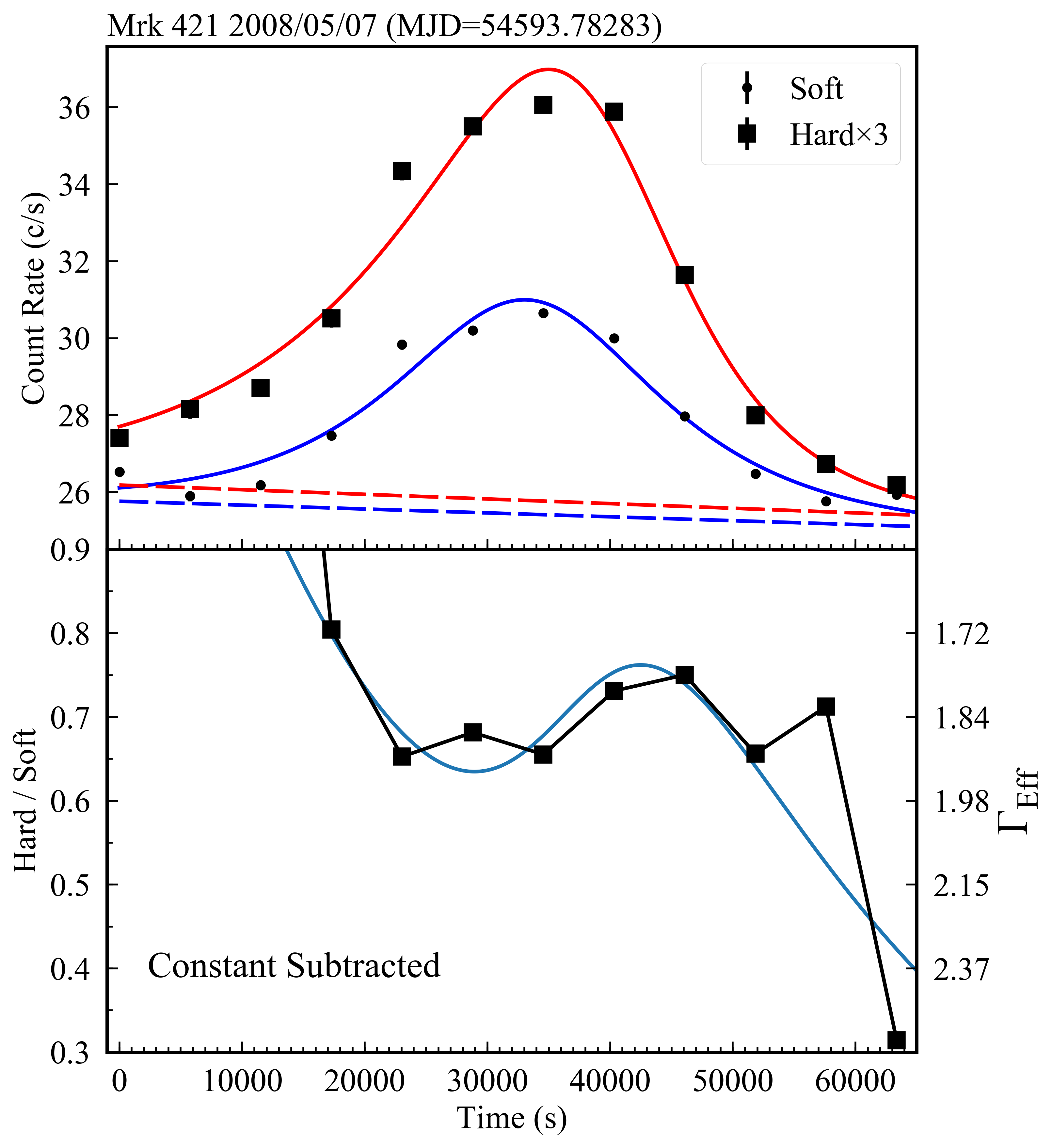}
\caption{An X-ray flare of Mrk~421 reported in \cite{Garson2010}.
    Top panel: The dots and squares are soft-band and hard-band lightcurves respectively \citep[cf. Fig.~6 of][]{Garson2010},
where the hard band has been multiplied by a factor of 3.
The dashed lines are the constant components in two bands.
In the bottom panel, the squares are the hardness ratios of the flare-only component.
The smooth curves are model lightcurves (top panel) or model predicted hardness ratios (bottom panel).}
\label{fig:mrk421lc}
\end{figure}

\begin{figure}[!htbp]
\figurenum{B2}
\includegraphics[width=0.48\textwidth]{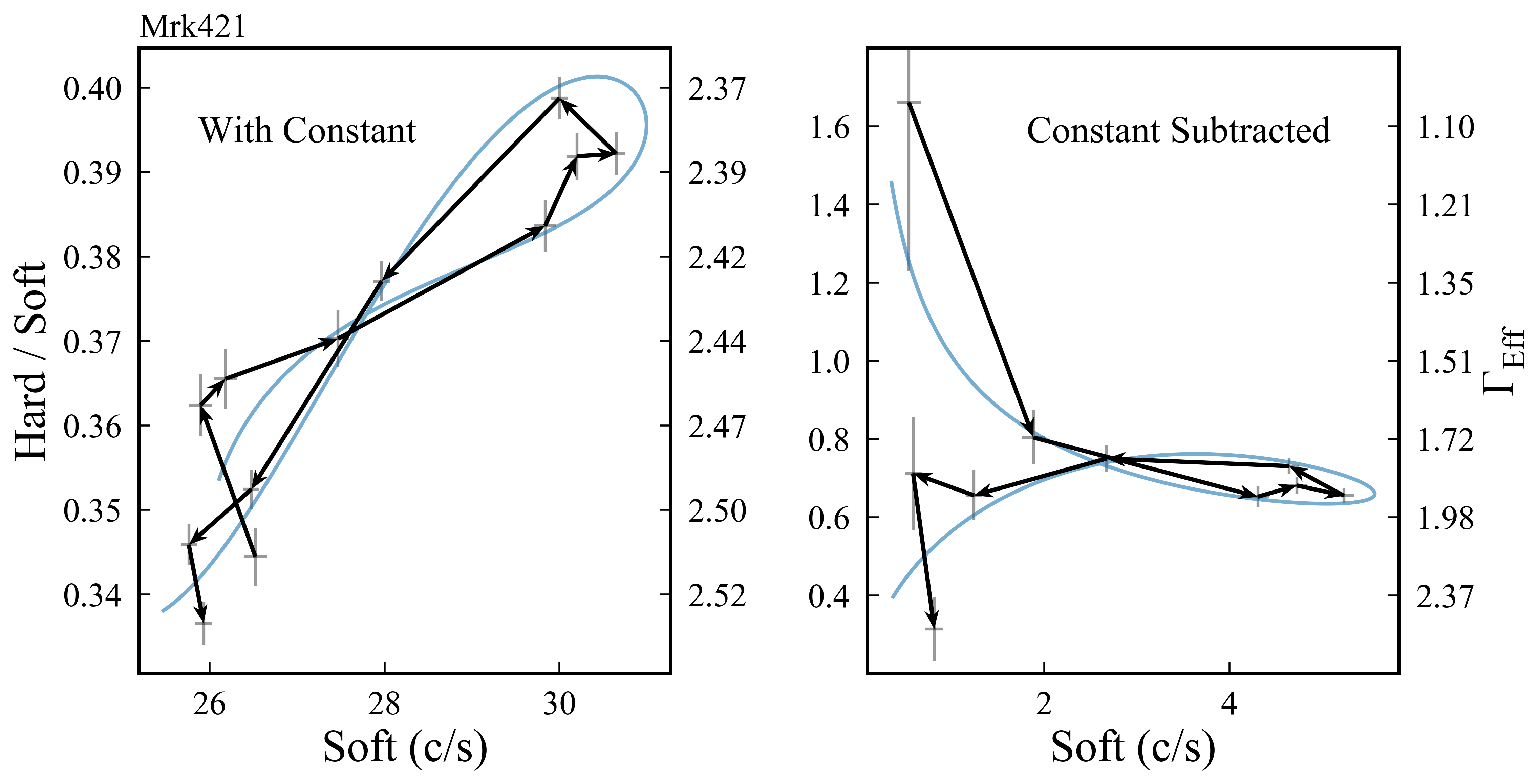}
\caption{Left: HR--flux diagram before the subtraction of the constant component \citep[cf. Fig.~6 of][]{Garson2010}.
The arrows indicate the direction of variation.
Right: HR--flux diagram after the subtraction of constant component.
Smooth curves are calculated from the model.
The effective photon index $\Gamma$ on the y-axis of right hand side is obtained using the response files of XIS, Galactic absorbed power-law model and a range of assumed photon indices.}
\label{fig:mrk421hr}
\end{figure}

The flare-only component of \pks{} shows an interesting spectral variability pattern,
but the noisy data prevent us from making strong conclusions.
The oblique ``8'' pattern in Fig.~\ref{fig:hr-flux_pks}(a) is reminiscent of
a flare of Mrk~421, which allows us to confirm our finding in another source using data from a different satellite.

We downloaded {\it Suzaku} data (ObsID 703043010) of Mrk~421 \citep{Garson2010}.
We only used data from the X-ray Imaging Spectrometer (XIS) onboard {\it Suzaku}.
The data were reprocessed and screened using {\sc Aepipeline} included in the {\it Suzaku} {\sc FTOOLS}.
In addition to the standard screening criteria, we also required the cutoff rigidity to be larger than 6 $\textrm{GV}/c$ following \cite{Garson2010}.
We extracted lightcurves separately from the cleaned event files of XIS0 and XIS3 using {\sc xselect} in initial 16 s bins.
The source region has an inner radius of 35 pixels and outer radius of 408 pixels, while the background region
is an annulus with an inner radius of 432 pixels and outer radius of 464 pixels.
The lightcurves were extracted in two energy bands: 0.5--2 keV and 2--10 keV, which are referred to as the soft band and hard band, respectively.
Lightcurves of the same energy band but different detectors (XIS0 and XIS3) were then merged.
We rebinned the lightcurves in a size of 5760 s in accordance with the orbital period of {\it Suzaku}.
Finally, the background lightcurves were subtracted to obtain the estimation of net count rates from the source.

We clipped the lightcurves and kept only the segment of {\it Flare 2} that is defined in \cite{Garson2010}.
We fitted the soft- and hard-band flares using a model that is analogous to Eq.~\ref{eqn:flare_profile}.
Only the constant component underlying the flare was replaced by a slowly-varying component using a linear function $F_{\rm c} = m + Slope\times t$.
This component is in long-term decline in both bands.
The fitted models were shown in the top panel of Fig.~\ref{fig:mrk421lc}.
The error bars of the lightcurves are exceedingly small, which renders the fitting
statistically unacceptable and suggests ultimate incorrectness of the model due to the existence of sub-structures.
However, variation of the smooth model curves in the figure match reasonably well with observational data.

The timescales are $\tau_{\rm r}=9630$ s and $\tau_{\rm d}=9261$ s in the soft band
and $\tau_{\rm r}=14756$ s and $\tau_{\rm d}=6796$ s in the hard band, which is consistent with the pattern that hard band rises slowly and decays fast.
The fitted long-term declining background was subtracted from the observed lightcurve of each
band to leave out the flare-only component.
We plotted hardness ratio variation with time and flux in the bottom panel of Fig.~\ref{fig:mrk421lc}
and the right panel of Fig.~\ref{fig:mrk421hr}.
Fig.~\ref{fig:mrk421hr} shows the same patterns as panels (a)(b) of Fig.~\ref{fig:hr-flux_pks}.
The effective photon indices ($\Gamma_\textrm{Eff}\sim 1.84$) are also close to the photon indices of the flaring components that are listed in Table~\ref{tab:spec_fit}.

\bibliographystyle{astron}
\bibliography{ms}
\end{document}